\documentclass[12pt]{iopart}

\usepackage{graphicx,color,rotating}
\newcommand{\op}[1]{\hat{#1}}
\newcommand{\ket}[1]{ \; \big| \; #1 \; \big>}

\newcommand{\bracket}[3]{ \;\big< \; #1 \; \big| \; #2 \;  \big| \; #3 \; \big>}

\begin{document}
\title[Bose-Fermi mixtures in 1D optical superlattices]{Bose-Fermi mixtures in 1D optical superlattices}
\author{Felix Schmitt, Markus Hild, and Robert Roth}
\address{Institut f{\"u}r Kernphysik, TU Darmstadt, Schlossgartenstra{\ss}e 9, 64289 Darmstadt, Germany
}
\begin{abstract}
The zero temperature phase diagram of binary boson-fermion mixtures in 
two-colour superlattices is investigated. The eigenvalue problem associated 
with the Bose-Fermi-Hubbard Hamiltonian is solved using an exact numerical 
diagonalization technique, supplemented by an adaptive basis truncation 
scheme. The physically motivated basis truncation allows to access larger 
systems in a fully controlled and very flexible framework. Several 
experimentally relevant observables, such as the matter-wave 
interference pattern and the condensate fraction, are investigated in order 
to explore the rich phase diagram. At symmetric half filling a phase similar 
to the Mott-insulating phase in a commensurate purely bosonic system is 
identified and an analogy to recent experiments is pointed out. Furthermore a phase of complete localization of the 
bosonic species generated by the repulsive boson-fermion interaction is 
identified. These localized condensates are of a different nature than the 
genuine Bose-Einstein condensates in optical lattices. 
\end{abstract}

\submitto{\JPB}
\maketitle

\section{Introduction}
Ultracold atomic gases in optical lattices have become a powerful and flexible 
experimental tool for studying strongly correlated quantum systems. The 
unique degree of experimental control makes them an ideal model system to 
investigate fundamental quantum phenomena which have an impact on several fields, 
ranging from condensed matter physics to quantum information processing. 
Theoretically these systems are well described in the language of the Hubbard 
model. For bosonic gases the Bose-Hubbard model \cite{Fischer} led to the 
prediction of a superfluid to Mott-insulator transition in ultracold atomic 
gases by Jaksch et al. \cite{Jaksch}. Since the experimental verification of 
this quantum phase transition by Greiner et al. \cite{Greiner} the field has 
developed rapidly.

Apart from the perfectly uniform lattice potentials generated by a single 
optical standing-wave, other lattice topologies have been discussed. Through 
the use of speckle patterns, optical lattices with random disorder have been 
realized \cite{Lye}. Already the superposition of two standing-wave 
lattices of different wavelengths leads to a superlattice with a spatial 
modulation of the lattice well depths. For all of these non-uniform lattice 
topologies, rich phase diagrams have been predicted 
\cite{scalettar,damski,rroth1}, which include new phases like an Anderson 
localized or a Bose-glass phase.

The technique of sympathetic cooling of fermionic atoms through thermal 
contact with a Bose gas has paved the way to experiments with Bose-Fermi 
mixtures in optical lattices. The phase diagram of these systems of mixed 
quantum statistics shows a multitude of new phenomena \cite{Albus, rroth2, 
Blatter, Mathey}. First experimental results \cite{Kenneth} using an 
admixture of a fermionic species to a Bose-Einstein condensate in an optical 
lattice reveal first effects of the inter-species interactions on the 
matter-wave interference pattern. In addition to the lattice topology, 
Feshbach resonances can be employed to tune the parameters of the 
Bose-Fermi-Hubbard Hamiltonian and thus to fully explore the zero-temperature 
phase diagram.

The aim of this paper is twofold. First, we investigate the effect of non-uniform 
lattice potentials---a simple two-colour superlattice in our case---on the 
phase diagram of binary Bose-Fermi mixtures. Second, we introduce a 
physically motivated basis truncation scheme to extend the calculations to experimentally 
relevant system sizes. The basic formulation of the model and benchmark 
results for purely bosonic systems are summarized in sections \ref{secBFH} 
and \ref{SINGLEB}, respectively. We use a direct numerical approach to 
solve the large-scale eigenproblem of the Bose-Fermi-Hubbard 
Hamiltonian, which constrains us to relatively small systems. In order to 
treat larger systems by direct diagonalization, we introduce an adaptive 
basis truncation scheme, which reduces the many-body basis to the physically 
relevant states.  The truncation scheme is discussed in section 
\ref{truncationscheme} and a first application to bosons in superlattices is 
presented in section \ref{secFINITE}. Finally, the phase diagram of 
Bose-Fermi-mixtures in superlattices is discussed in detail in sections 
\ref{BFMIX} and \ref{secBFTRUNC}.

\section{Bose-Fermi Hubbard Hamiltonian with a superlattice potential}\label{secBFH}

We consider a system of $N_a$ bosonic and $N_c$ fermionic atoms in a one-dimensional optical lattice with $I$ lattice sites at zero temperature. A detailed investigation of these systems in regular lattice potentials can be found in Refs. \cite{rroth2, Mathey}. In extension of these studies, we are going to discuss the impact of spatial lattice modulations on the phase diagram of boson-fermion mixtures. In the simplest case we assume a two-colour superlattice potential, generated by a superposition of two standing-wave lattices of different wavelengths. Provided the lattice is sufficiently deep, i.e., the gap between the ground and first excited band is sufficiently large, we can restrict ourselves to the lowest energy band \cite{Zwerger}. A straightforward generalization of the single-band Bose-Hubbard Hamiltonian leads to the single-band Bose-Fermi-Hubbard Hamiltonian. In second quantisation it reads \cite{rroth1, Albus, Blatter, Krauth}:
\begin{eqnarray}\label{BFH}
  \op{H} 
  &=& - \, J \, \sum_{l=1}^{I} \; \left( \; \op{a}_{l+1}^{\dagger} \op{a}_l^{} +
    \op{c}_{l+1}^{\dagger} \op{c}_l^{}  + \; h.a. \right)
    + \,\frac{1}{2} \, V_{aa}  \, \sum_{l=1}^{I} \op{n}_l^{(a)} \left(\op{n}_l^{(a)}-1 \right) \nonumber\\
	&+& \,  V_{ac} \, \sum_{l=1}^{I}  \; \op{n}_l^{(a)} \; \op{n}_l^{(c)}
	+\Delta \sum_{l=1}^{I} \epsilon_l \,  \left(\op{n}_l^{(a)} + \op{n}_l^{(c)} \right).
\end{eqnarray}
The operators $\op{a}^{\dagger}_l$ ($\op{a}^{}_l$) create (annihilate) a boson in the lowest Wannier state localized at lattice site $l$. The operators $\op{c}^{\dagger}_l$ ($\op{c}^{}_l$) are the corresponding fermion creation (annihilation) operators, all fulfilling their particular commutation or anticommutation relations. The occupation number operators for the bosonic and fermionic component are given by $\op{n}_l^{(a)} = \op{a}_{l}^{\dagger} \op{a}_l^{}$ and $\op{n}_l^{(c)} =\op{c}_{l}^{\dagger} \op{c}_l^{}$, respectively.

The first term of the Hamiltonian (\ref{BFH}) describes the tunnelling of atoms between adjacent lattice sites with a tunnelling matrix element $J$. We assume periodic boundary conditions, thus tunnelling between sites $I$ and $1$ is included. The next two terms describe the boson-boson and the boson-fermion on-site interaction with strengths $V_{aa}$ and $V_{ac}$, respectively. Due to the single-band description and the Pauli-principle there is no on-site fermion-fermion interaction. The last term describes an on-site potential energy in which $\Delta$ is the maximum amplitude and the $\epsilon_l$ specify the superlattice topology. The parameters of the Hubbard Hamiltonian encapsulate the properties of the many-body system, such as the depth and shape of the lattice potential and the scattering lengths of the boson-boson and the boson-fermion interaction. They are defined via matrix elements in the localized Wannier states. The presence of a superlattice potential is reflected mainly in the site-dependent single-particle energies $\epsilon_l$. The site-dependence of $J$, $V_{aa}$ and $V_{ac}$ can be neglected in good approximation.

The many-body states within the Hubbard model can be conveniently represented in an occupation number basis. For a mixed boson-fermion system, the full occupation number basis is constructed via a direct product of the bosonic and the fermionic occupation number bases. A general many-body state can then be written as a superposition of all possible combinations of the number states:
\begin{equation}\label{BFGS}
  \ket{\psi} 
  = \sum_{\alpha=1}^{D_a} \sum_{\beta=1}^{D_c} 
     C_{\alpha \beta} \ket{ \{ n_1^{(a)}, ... \, ,n_I^{(a)} \}_{\alpha}} \otimes 
     \ket{ \{ n_1^{(c)},...\, , \, n_I^{(c)} \}_{\beta} } \;.
\end{equation}
The sets of occupation numbers $\{n_1^{(a)}, ... \, ,n_I^{(a)}\}$ for the bosonic component comprise all possible distributions of the $N_a$ atoms over the $I$ lattice sites with $\sum_{l}n_l^{(a)} = N_a$. For the fermionic occupation numbers $\{ n_1^{(c)},...\, , \, n_I^{(c)} \}$ the additional constraint $n_l^{(c)}\in\{0,1\}$ reflects the Pauli principle. The basis dimension grows rapidly with increasing lattice size and particle numbers, the dimensions of the bosonic and the fermionic number bases are given by $D_a=(N_a+I-1)!/(N_a!(I-1)!)$ and $D_c=I!/(N_c!(I-N_c)!)$, respectively, with the total dimension being the product of the two sub-space dimensions. 

Because of the factorial growth of the basis dimension, any exact solution of the many-body problem using the full basis is restricted to moderate system sizes. To solve the eigenvalue problem of the Bose-Fermi-Hubbard Hamiltonian, we are using refined Lanczos-type algorithms \cite{Arpack}, which enable us to treat basis dimensions of the order of $10^6$. 

One of the major advantages of exact diagonalization techniques is their applicability throughout the whole phase diagram, from the coherent superfluid regime to the strongly correlated Mott-insulator \cite{Jaksch, rroth1}. The price for this universality is the restriction to moderate system sizes due to the huge basis dimensions. Our aim is to increase the computationally feasible system sizes by a physically motivated, adaptive truncation scheme, which selects the important basis states for a certain region of the phase diagram out of the complete basis. This allows us to approach the experimentally relevant system sizes, to check for finite size effects, and to use the truncated basis in time-dependent simulations \cite{markus}.

\section{Single component Bose gas -- a reminder}\label{SINGLEB}

\begin{figure}[]
\begin{indented}\item
\footnotesize
\includegraphics[]{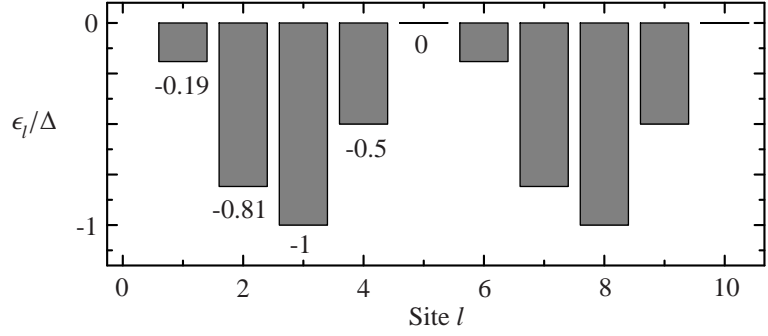}
\end{indented}
\caption{Topology of the two-colour superlattice with 2 super-cells. $\Delta$ is the maximum amplitude of the superlattice potential. The $\epsilon_l$ define its topology.\label{SL}}
\end{figure}

We start with a brief reminder of some previous results \cite{rroth1} on Bose gases in superlattices, which later on will serve as a benchmark for the accuracy of truncated calculations. Consider a single-component Bose gas with $N_a=10$ particles in a two-colour superlattice with $I=10$ lattice sites. The topology of the superlattice is defined by the set of on-site energies $\epsilon_l$ depicted in figure \ref{SL}. The Hamiltonian follows from (\ref{BFH}) by omitting all fermionic contributions. Expressing all energies in units of the tunnelling energy $J$, the interaction strength $V_{aa}$ and the superlattice amplitude $\Delta$ are the remaining two parameters.

\begin{figure}
\begin{center}
\includegraphics[width=1\textwidth]{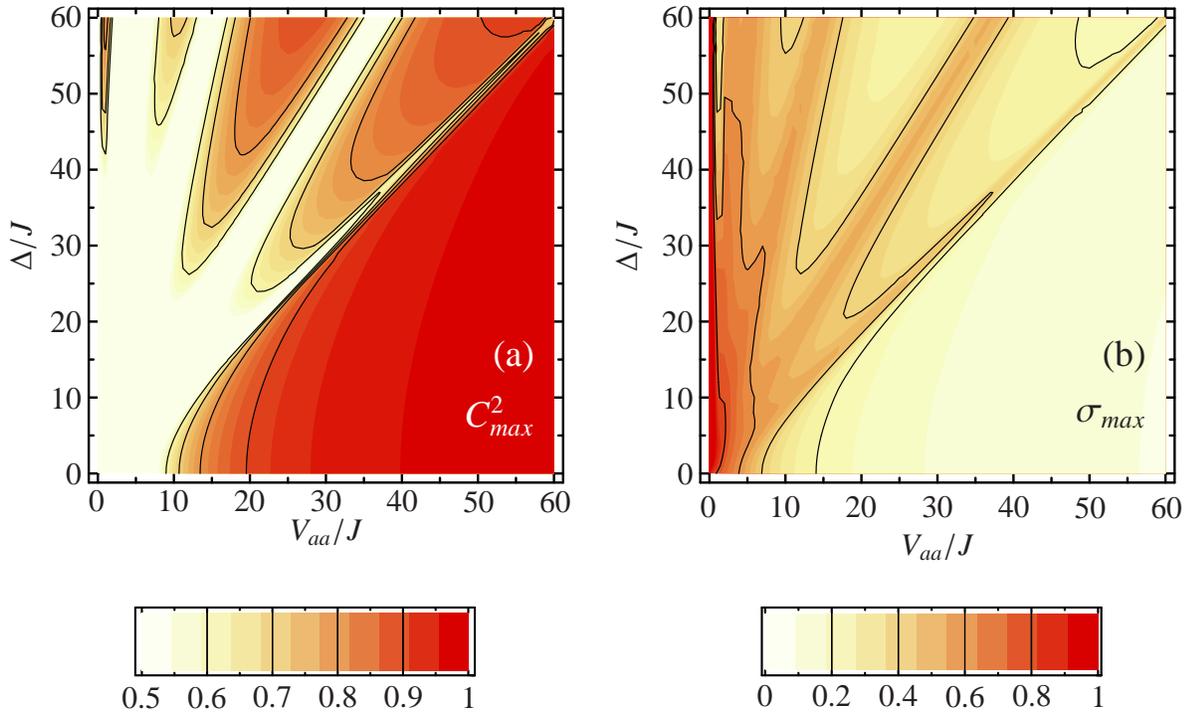}
\end{center}
\caption{(a) Maximum coefficient $C^2_{max}$ in the expansion of the number states (\ref{BFGS}) in the $V_{aa}-\Delta$ plane. (b) Maximum number fluctuations $\sigma_{max}$. Both for $N_a=10$ bosonic particles and $I=10$ lattice sites. \label{CMAX1}}
\end{figure}

A first impression of the rich structure of the phase-diagram in the $V_{aa}-\Delta$ plane is provided by the maximum coefficient in the expansion (\ref{BFGS}) of the groundstate:
\begin{equation}\label{Cmax}
C^{2}_{max}=\mathrm{max} \left\{ C^2_{\alpha} \right\} \;
\end{equation}
depicted in figure \ref{CMAX1}(a). Note that the index $\beta$ has been dropped due to the absence of the fermionic species. In the case of small values of $C^2_{max}$ there are no preferred number states. For small superlattice amplitudes $\Delta$ and boson-boson interaction strengths $V_{aa}$, the system is in the superfluid phase, where virtually all number states contribute. If $C^2_{max}$ is close to $1$, the groundstate is dominated by a single number state. In the Mott-insulator phase at large interaction strengths $V_{aa}$ this is the number state with one particle per lattice site. Others are suppressed because of their interaction energy contribution. The homogenous Mott-insulating phase is present as long as the superlattice amplitude is smaller than the interaction energy $\Delta < V_{aa}$ as can be seen in figure \ref{CMAX1}(a). As soon as they are approximately equal, $\Delta \approx V_{aa}$, double occupations of the deepest lattice wells become energetically possible. In this transition region many number states contribute, giving rise to the reduced $C^2_{max}$ at $\Delta \approx V_{aa}$. This is the onset of the quasi Bose-glass phase, where the interplay of lattice irregularities and interaction causes redistributions of the particles within the system. These redistributions of the particles also manifest in increasing number fluctuations
\begin{equation}\label{fluc}
\sigma^2_{max}=\mathrm{max} \left\{\bracket{\psi}{\op{n}^2_l}{\psi}-\bracket{\psi}{\op{n}_l^{}}{\psi}^2 \right\}
\end{equation}
shown in figure \ref{CMAX1}(b). Throughout the phase diagram one can clearly identify the complementary behaviour of  $C^2_{max}$ and the number fluctuations $\sigma_{max}$. The lobes result from the finite differences of the on-site energies in the superlattice potential (compare figure \ref{SL}). In the limit of an infinite lattice with random disorder these stable regions will vanish and the genuine Bose-glass phase will emerge. Another characteristic regime is the localized phase in which all particles are gathered at the deepest lattice well of each super-cell. It is obvious that this phase only exists at very small interaction strengths. The effect of localization due to disorder in the lattice is called \textit{Anderson localization} \cite{Gebhard}.

\section{Basis truncation scheme}
\label{truncationscheme}

One possible way to extend these direct solutions of the eigenvalue problem to larger systems consists in a controlled and physically motivated truncation of the underlying basis. The challenge is to decide which number states are important and which can be discarded from the outset. As an example consider a system deep within the Mott-insulating phase. Number states with several particles at one particular lattice site are associated with very large interaction energies and their contribution to the ground state is negligible. Those less important number states can be omitted from the beginning, leading to a significant reduction of the basis dimension.

A simple but very efficient a priori measure for the importance of individual number states is the expectation value of the Hamiltonian. The following inequality is used to select the relevant basis states:
\begin{equation}\label{TRUNC}
 \bracket{ _{\alpha} \{ n_1,...,n_I \}}{\op{H}}{ \{ n_1,...,n_I \}_{\alpha}}\leq E_{trunc}.
\end{equation}
Hence, number states associated with interaction or on-site energy contributions above a specific truncation energy $E_{trunc}$ are discarded. By varying the parameters $V_{aa}$, $V_{ac}$, and $\Delta$ one can optimize the basis for a particular region of the phase diagram. By varying the truncation energy $E_{trunc}$ one can adjust the size of the remaining basis and at the same time assess the truncation errors. The truncation scheme works particularly well in regions where the ground state is composed of a few dominant basis states, i.e. in regions of large $C^2_{max}$. On the other hand, in regions where almost all number states contribute significantly, e.g. in the superfluid phase, the truncation scheme will be less reliable.

The generation of the truncated basis starts from a reference number state that minimizes the energy expectation value. With respect to this reference state, multiple particle-hole excitations are generated until the new states violate the inequality (\ref{TRUNC}). If the truncation energy $E_{trunc}$ is large enough, this process generates the complete set of number states.

As a first benchmark, we study the effect of the basis truncation for the single-component Bose gas discussed earlier. For the sake of clarity and simplicity the interaction energy is fixed at $V_{aa}/J=30$. The calculations performed with the complete basis are compared to calculations with two different truncation levels. The phase diagram is split into three parts, each of them is calculated with an optimized basis. Table \ref{tabular} lists the different ranges in $\Delta/J$ together with the corresponding basis dimensions and reference states.

\begin{table}
\small
\begin{center}
\begin{tabular}{|l|c|c|c|}
\hline
system & sections ($\Delta/J$) & dimension & reference number state\\
\hline\hline
strong truncation &  1-28 & 91 & $\ket{\{1,1,1,1,1,1,1,1,1,1\}}$\\ \cline{2-4}
$\approx 0.3 \%$ of complete basis & 29-50 &  288 & $\ket{\{1,1,2,1,0,1,1,2,1,0\}}$\\ \cline{2-4}
 & 51-150 & 199 & $\ket{\{0,2,2,1,0,0,2,2,1,0\}}$ \\ 
\hline\hline
moderate truncation &  1-29 & 3743 & $\ket{\{1,1,1,1,1,1,1,1,1,1\}}$\\ \cline{2-4}
$\approx 4 \%$  of complete basis & 30-50 & 1866 & $\ket{\{1,1,2,1,0,1,1,2,1,0\}}$\\ \cline{2-4}
& 51-150 & 2737 & $\ket{\{0,2,2,1,0,0,2,2,1,0\}}$ \\ 
\hline
complete basis & 1-150 & 92378 &--\\
\hline
\end{tabular}
\end{center}
\caption{Basis dimension and reference states for different truncation levels and different ranges of $\Delta/J$.\label{tabular}}
\end{table}

Since this approach fulfils the variational principle, we allow for overlaps at the borders of different sections and select the basis that yields the lowest groundstate energy. The strong truncation includes 3-particle--3-hole excitations with respect to an appropriate reference state chosen for the particular region in the phase diagram. The remaining dimension is about 0.3\% of the complete basis. For the basis with moderate truncation, higher-order excitations are also included. Nevertheless, the dimensions of the moderate truncated basis is still reduced to 4\% of the complete basis.

\begin{figure}
\includegraphics[width=1\textwidth]{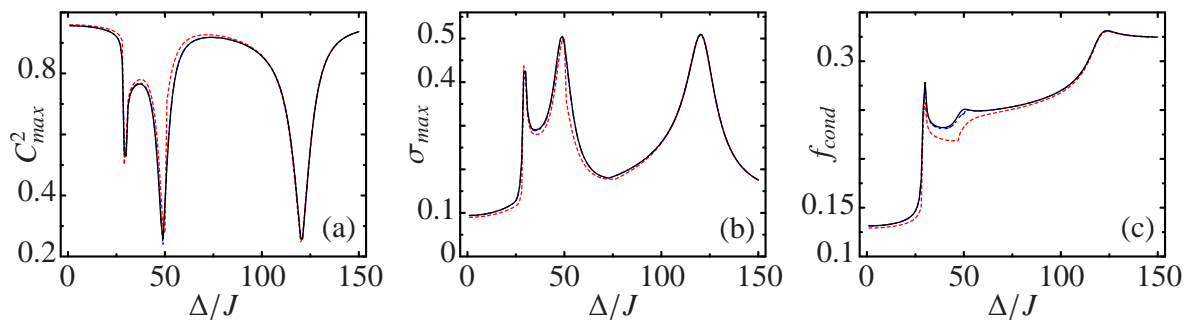}
\caption{$N_{a}=10$ bosonic particles on $I=10$ lattice sites at fixed interaction strength $V_{aa}/J=30$. (a) maximum coefficient $C^2_{max}$, (b) maximum number fluctuation $\sigma_{max}$, (c) condensate fraction $f_{cond}$, all plotted over the amplitude of the superlattice $\Delta/J$. Calculations within the complete basis (solid line), truncated to 4\% of the complete basis (dashed-dotted, blue online), and truncated to 0.3\% of the complete basis (dashed, red online). The results of the full basis and the truncated basis with 4 \% of the full dimension coincide almost everywhere.\label{OBS1}}
\end{figure}

The dependence of the maximum coefficient $C^2_{max}$ on the amplitude of the superlattice $\Delta$ at fixed interaction strength is depicted in figure \ref{OBS1}(a). If the amplitude $\Delta$ is smaller than the interaction strength $V_{aa}$, the system is in the homogenous Mott-insulator phase and the number state with one particle per lattice site dominates. Further increase of the superlattice amplitude leads to the onset of the quasi Bose-glass phase, where a pattern of regions with small values of $C_{max}^2$ and large number fluctuations and regions with small number fluctuations and large values of $C_{max}^2$ emerges as depicted in panels (a) and (b) of figure \ref{OBS1}. The differences between the calculations with the full and the truncated basis are barely visible. Only for the strongly truncated basis with 0.3\% of the full basis dimension, noticeable deviations occur. The basis using a moderate truncation to 4\% of the full dimension yields results which are practically identical to the full calculation. 

A well-defined observable is the fraction of particles that undergo Bose-Einstein condensation. The largest eigenvalue $\lambda_1$ of the one-body density-matrix defined by:
\begin{equation}\label{obdm}
\rho^{(1)}_{ll'}=\bracket{\psi}{\op{a}^{\dagger}_{l'} \op{a}^{}_{l} }{\psi}
\end{equation}
is used to calculate the condensate fraction \cite{rroth1, Penrose}:
\begin{equation}\label{condfrac}
f_{cond}=\frac{\lambda_1}{N_{a}}.
\end{equation}
The eigenvectors of the one-body density-matrix are called \textit{natural orbitals} --- in a regular lattice these are Bloch functions.

Figure \ref{OBS1}(c) illustrates the dependence of the condensate fraction $f_{cond}$ on the superlattice amplitude $\Delta$ for fixed $V_{aa}/J=30$. Leaving the homogenous Mott-insulator phase, the condensate fraction increases rapidly, reaching a local maximum at the transition point to the quasi Bose-glass phase at $\Delta/J=30$. After a slight decrease within the first lobe of the Bose-glass phase it increases further. It seems that in the limit of large superlattice amplitudes $\Delta$, two separate ``localized condensates'' \cite{pugatch} form at the deepest lattice wells. The term condensate has to be used with caution in this context -- we will come back to this point in section \ref{BFMIX}.

Again, the deviations between truncated and full calculations are generally small. At the local maximum at $\Delta/J=30$ the exact calculation yields $f_{cond}=0.278$, for the moderate truncation $f_{cond}=0.276$ and for the strong truncation $f_{cond}=0.258$. It is an intrinsic feature of the truncation to underestimate the condensate fraction because of the number states that are missed during the calculation of the one-body density-matrix elements. The kink in figure \ref{OBS1}(c) at $\Delta/J=50$ (dashed line) is due to the change of the basis at this point (see table \ref{tabular}). In principle one could adopt another optimized basis in this region in order to improve the description. 

\begin{figure}
\includegraphics[width=1\textwidth]{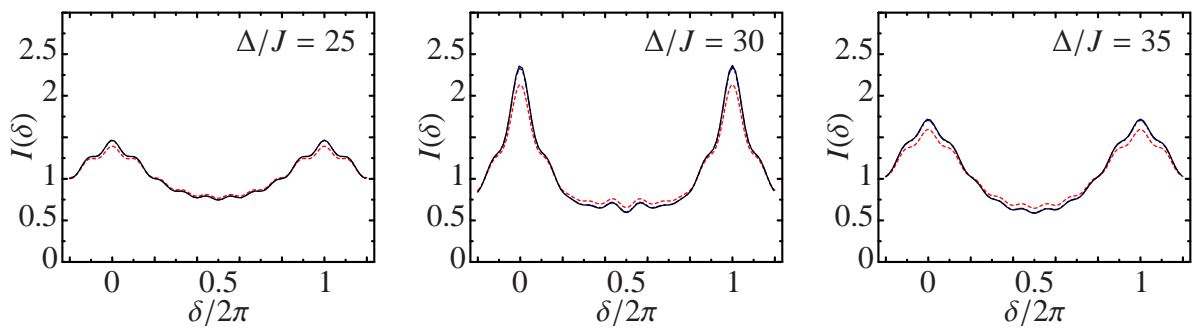}
\caption{Matter-wave interference pattern in the vicinity of the Mott-insulator to the quasi Bose-glass transition point. Calculations within the complete basis (solid line), truncated to 4\% of the complete basis (dashed-dotted, blue online), and truncated to 0.3\% of the complete basis (dashed, red online). The dashed-dotted line is hardly visible because of the good reproduction of the complete calculation.\label{PATTERN1}}
\end{figure}

An observable that is directly accessible to the experiment is the matter-wave interference pattern after release from the lattice and free expansion. After switching off the confining potential, the intensity of the matter-wave at a point $x$ can be written as \cite{rroth1}
\begin{equation}
	\mathcal{I}(x)= \bracket{\psi}{\op{A}^{\dagger}(x) \,\op{A}(x)}{\psi} \;.
\end{equation}
Since we are not interested in the spatial envelope of the interference pattern, the amplitude operator $\op{A}(x)$ depends only on the phase $\phi_l(x)$ acquired on the way from site $l$ to the observation point $x$,
\begin{equation}
	\op{A}(x) = \frac{1}{\sqrt{I}} \sum_{l=1}^{I} e^{\mathbf{i} \, \phi_l(x)} \, \op{a}_l^{}\;.
\end{equation}
Considering microscopic distances between the lattice sites and a macroscopic distance from the lattice to the observation point, we assume a constant phase difference between adjacent sites $\delta=\phi_{l+1}(x)-\phi_l(x)$. This far-field limit leads to the following expression for the matter-wave intensity:
\begin{equation}\label{matpat}
	\mathcal{I}(\delta)=\frac{1}{I} \sum^{I}_{l,l'=1} e^{\mathbf{i}(l-l')\delta} \, \bracket{\psi_0}{\op{a}^{\dagger}_{l'} \op{a}_l^{}}{\psi_0}= \frac{1}{I} \sum^{I}_{l,l'=1} e^{\mathbf{i}(l-l')\delta} \; \rho^{(1)}_{ll'} \;,
\end{equation}
in which we used equation (\ref{obdm}). This expression is closely related to the quasi-momentum distribution. If $\delta$ is equal to one of the possible quasi-momenta $q_k=2 \pi k /I$, where $k=0,1, \ldots ,I-1$, we obtain the respective quasi-momentum occupation numbers. Exemplarily figure \ref{PATTERN1} shows the matter-wave interference pattern for three values of the superlattice amplitude $\Delta$ in the vicinity of the transition point from the homogenous Mott-insulator to the quasi Bose-glass phase. From the sequence of interference patterns one can see that the height of the peak at quasi-momentum $q=0$ reaches a maximum at $\Delta=V_{aa}$ with an absolute value smaller than the condensate occupation number. Note that in a regular lattice the condensate wave function---the lowest natural orbital---is identical to the quasi-momentum zero Bloch function. Hence, in that case the occupation of the quasi-momentum zero mode is equal to the lowest eigenvalue of the one-body density matrix. This relation does not hold for irregular lattice potentials.

As already mentioned, the truncation scheme is expected to deviate most from the complete calculations if a large number of basis states is necessary to represent the ground state. But even in the transition region, where this occurs, less than 1\% of the complete set of number states is sufficient to reproduce the properties of the system qualitatively. The basis including 4\% of the complete set of number states included shows virtually no deviation from the exact calculation. It seems that the estimate (\ref{TRUNC}) is a suitable a priori measure to select the physically relevant number states. It is obvious that the more irregular the optical lattice is, the less energetically similar number states exist and the better the truncation scheme works.

\section{Multiple super-cells and finite size effects}\label{secFINITE}

As a first application of the truncated basis in a regime where calculations with the full basis are not feasible anymore, we study the impact of finite size effects for a single-component Bose gas in a superlattice. To this end, we extend the two-colour lattice to three and four super-cells, corresponding to a problem of 15 bosons on 15 lattice sites ($N_a=I=15$) and 20 bosons on 20 lattice sites ($N_a=I=20$), respectively. We employ truncations to 1\% and 0.001\% of the respective full basis dimensions. 

\begin{figure}
\begin{indented}\item
\includegraphics[width=.8\textwidth]{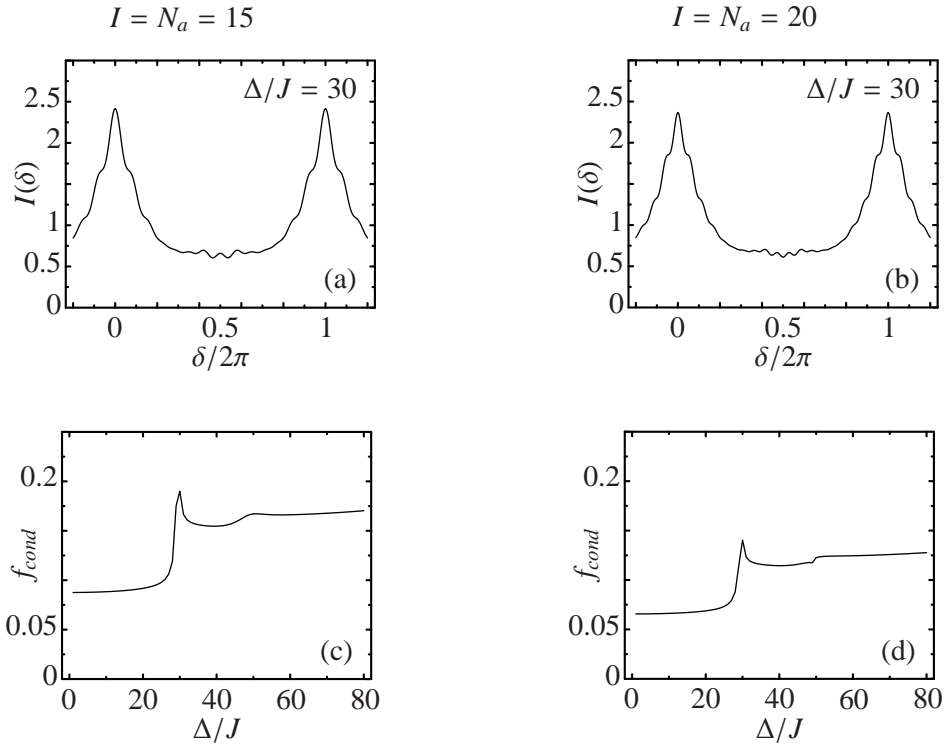}
\end{indented}
\caption{Left: $N_{a}=I=15$ system with 3 super-cells, right: $N_{a}=I=20$ system with 4 super-cells. (a) and (b): Matter-wave interference pattern $\mathcal{I}(\delta)$ at the homogenous Mott-insulator to quasi Bose-glass transition point $V_{aa}/J=\Delta/J=30$. (c) and (d): condensate fraction $f_{cond}$ over the superlattice amplitude $\Delta$.\label{PATTERN2}}
\end{figure}

The resulting matter wave interference patterns for $V_{aa}/J=\Delta/J=30$, i.e., at the transition from homogeneous Mott-insulator to the quasi Bose-glass phase, are depicted in the upper part of figure \ref{PATTERN2}. They can be compared to the result of the full calculation for a system of two super-cells shown in figure \ref{PATTERN1}. The matter-wave interference patterns seem to be barely affected by the truncation and are of the same general shape for all system sizes. Only the increasing number of possible quasimomenta $q_k=2\pi/I$ with $k=1, 2, \ldots I-1$ is reflected in the substructure of the interference peaks.

Figure \ref{PATTERN2} also demonstrates the dependence of the condensate fraction on $\Delta/J$ at fixed $V_{aa}/J=30$ for the truncated three and four super-cell system. The comparison with the complete calculation for the two super-cell system displayed in figure \ref{CMAX1} reveals that the condensate fraction is generally reduced. The reason for this is two-fold: First, the basis truncation leads to an underestimation of the condensate fraction as discussed in section \ref{truncationscheme}. Second, finite size effects are relevant for the condensate fraction. The simplest indication for this results from the existence of a lower bound for $f_{cond}$ which is proportional to $1/I$. The largest eigenvalue $\lambda_1$ of the one-body density matrix (\ref{condfrac}), which defines the condensate fraction, has a lower bound which results from the normalization of the one-body density matrix: 
\begin{equation}
N_{a} 
= \tr(\rho^{(1)}_{ll'})
= \sum_{i=1}^I \lambda_i \leq \sum_{i=1}^I \lambda_1
= I \lambda_1  
\quad \Rightarrow \quad 
\lambda_1 \geq \frac{N_a}{I} \;.
\end{equation}
Hence, the smallest possible value for the condensate fraction is given by $f_{cond}^{min}=1/I$. For the genuine Mott-insulating phase in a regular lattice of increasing size the condensate fraction would go to zero like $1/I$. In agreement with this intrinsic size dependence, the condensate fraction within the Mott-insulating phase shown in figure \ref{PATTERN2} tends to smaller values.

From these investigations we can conclude, that for the major observables finite size effects are under control and that calculations performed in small systems are well suited for qualitative and even semi-quantitative predictions for larger systems.

\section{Binary Bose-Fermi Mixtures}
\label{BFMIX}

Recent developments in the sympathetic cooling of boson-fermion mixtures in optical lattices pave the way to experimental studies of dilute, degenerate gases with mixed quantum statistics \cite{Kenneth}.
Various aspects of boson-fermion mixtures in uniform lattices have been discussed before \cite{lewenstein,rroth2}. In this section, we study the effect of an additional superlattice potential on the phase diagram. 

The following calculations were performed for a system with $I=10$ lattice sites and $N_a=N_c=5$ particles of each species using the complete basis of $D_a \cdot D_c=504504$ number states. The boson-boson interaction energy is fixed to $V_{aa}/J=20$ in order to simplify the discussion --- other (non-zero) values would not change the basic structure of the phase diagram but only cause a rescaling. 

\begin{figure}
\includegraphics[width=1\textwidth]{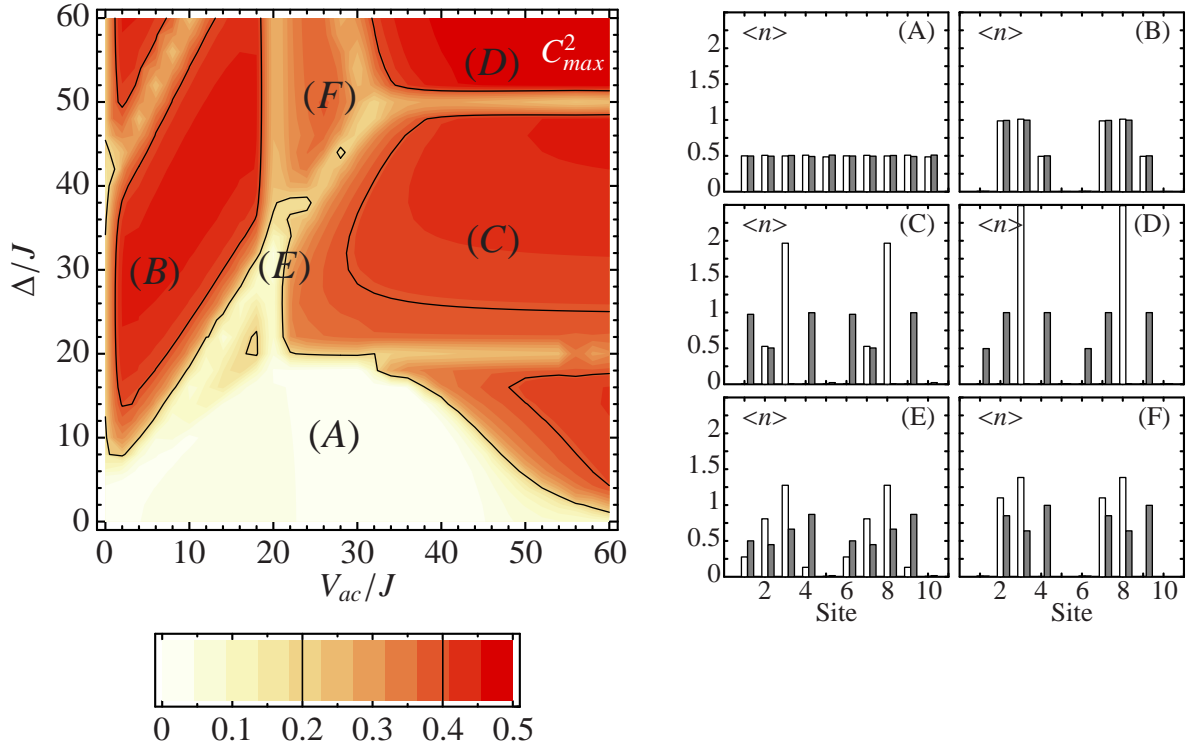}
\caption{Left panel: $C_{max}^2$ of a binary Bose-Fermi mixture with $N_a=N_c=5$ particles and $I=10$ lattice sites. Right panel: The mean occupation numbers (white for bosons and gray for fermions) for the selected points in the phase diagram. The boson-boson interaction energy is fixed at $V_{aa}/J=20$.\label{CMAX2}}
\end{figure}

Figure \ref{CMAX2} shows the maximum coefficient $C_{max}^2$ as a function of the boson-fermion interaction strength $V_{ac}/J$ and superlattice amplitude $\Delta/J$. Already the maximum coefficient of the number state representation of the ground state reveals the rich structure of the phase diagram. We can identify several straight lines or valleys of minimum $C_{max}^2$ that subdivide the phase diagram into different regions. There is a horizontal line at $\Delta= V_{aa}$ and a vertical line at $V_{ac}=V_{aa}$, whose locations are determined by the boson-boson interaction strength and which are independent of the precise actual lattice topology. In contrast, the additional diagonal valleys are governed by the details of the lattice topology, i.e., the set of on-site energies $\epsilon_{l}$.

Typical occupation number distributions for bosons and fermions in the different areas of the phase diagram separated by those valleys are also presented in figure \ref{CMAX2}. They correspond to the values of the boson-fermion interaction strength and the superlattice amplitude marked by the labels in the phase-diagram. The characteristics of the different regions can be summarized as follows: 

\begin{description}
\item[(A)] In this region, the atoms are uniformly distributed over the lattice because of the dominant repulsive interactions. All number states with few particles per site are favoured since the strong boson-fermion repulsion suppresses states with different species occupying the same site. Because of the commensurate filling and the strongly repulsive interactions between bosons as well as between bosons and fermions this phase is closely related to the Mott-insulating phase in a purely bosonic system.

\item[(B)] The mean occupation numbers of both species follow the shape of the superlattice potential. In the upper-left corner of this region two bosons are able to share the deepest superlattice wells with one fermion. Increasing boson-fermion repulsion $V_{ac}$ forces one boson to leave the deepest superlattice well and the mean occupation numbers exhibit a profile as depicted in inset (B).

\item[(C)] Above $\Delta/J=20=V_{aa}/J$ the on-site energy of the deepest superlattice wells exceeds the boson-boson interaction energy and they are capable to hold 2 bosonic particles. This double occupation occurs if the boson-fermion interaction prevails the boson-boson interaction and forces the fermion to leave the deepest well. 

\item[(D)] Complete localization of the bosons at the deepest superlattice sites induced by the strong repulsive boson-fermion interaction despite of the boson-boson repulsion. In a purely bosonic system with $V_{aa}/J=20$, a full localization would occur only for extremely strong superlattice potentials with amplitudes $\Delta/J>210$. 
The presence of the fermions and the strong boson-fermion repulsion, causing a separation of bosons and fermions, promote the localization. The borders of this region of separation induced localization depend not only on the interaction strengths and the superlattice amplitude, but also on the filling. The more particles per super-cell are available, the further the onset of localization is pushed to larger superlattice amplitudes.

\item[(E)] Transition region between (B) and (C). This area of subtle rearrangements is governed by the finite differences of the on-site energies $\epsilon_l$ of the superlattice potential.

\item[(F)] Transition region between (B) and (D). The fermions on sites $l=2$ and 7 successively force the bosons to the state of complete localization. If the bosons were the only species this would happen at much larger superlattice amplitudes.

\end{description}

\begin{figure}
\includegraphics[width=1\textwidth]{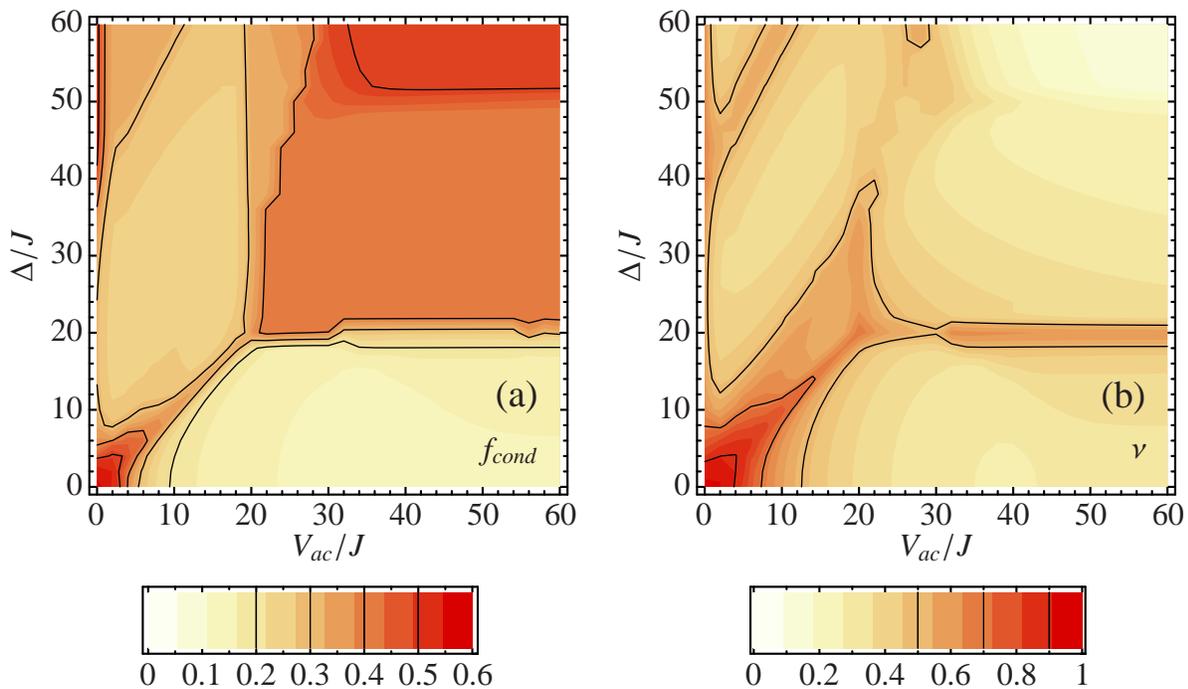}
\caption{Binary Bose-Fermi mixture with $N_a=N_c=5$ particles on $I=10$ lattice sites at fixed boson-boson interaction energy $V_{aa}/J=20$. (a) condensate fraction $f_{cond}$ of the bosonic species. (b) fringe visibility $\nu$ of the bosonic species. Although the condensate fraction at the top right is large, the fringe visibility vanishes there. This is due to the different nature of the localized condensates (see text).\label{COND2}}
\end{figure}

The phase diagram in terms of the condensate fraction $f_{cond}$ of the bosonic species is depicted in figure \ref{COND2}(a). Only three areas with a significant condensate fraction are visible. In the lower left corner---at small boson-fermion interaction strengths and superlattice amplitudes---large condensate fractions occur despite of the strongly repulsive boson-boson interaction ($V_{aa}/J=20$), since the incommensurate boson filling averts a Mott-insulator phase. The condensate is depleted as soon as either the increasing boson-fermion interaction leads to a Mott-like phase, or the superlattice tends to localize the particles. The large condensate fractions in an extended region in the upper right part and along the upper part of the $V_{ac}/J=0$ axis have to be treated with care and are of the same character as the localized condensates discussed in section \ref{truncationscheme}. The almost complete localization of the bosonic particles at the deepest lattice sites $l=3$ and 8 manifests in the corresponding diagonal matrix elements, i.e., the occupation numbers, of the one-body density matrix $\rho^{(1)}_{33}$ and $\rho^{(1)}_{88}$. In comparison, the other matrix elements are small except for the off-diagonal elements $\rho^{(1)}_{38}$ and $\rho^{(1)}_{83}$. This results in two large eigenvalues of the one-body density matrix and, therefore, in our description, in a large condensate fraction. Note, however, that this `condensate' is completely different from the condensate associated with the superfluid phase in a homogenous lattice, where all particles are delocalized and exhibit long-range phase coherence. This characterization of the condensate is closely related to the existence of off-diagonal long-range order \cite{Penrose} in the one-body density matrix, which is absent in the case of the localized condensates. 

This interpretation is also supported by the behaviour of the fringe visibility $\nu$ shown in figure \ref{COND2}(b). The visibility is directly derived from the matter-wave interference pattern (\ref{matpat}) and defined as:
\begin{equation}\label{nu}
	\nu = \frac{{\mathcal I}_{max}-{\mathcal I}_{min}}{{\mathcal I}_{max}+{\mathcal I}_{min}}.
\end{equation}
Clearly, the visibility is large at small boson-fermion interaction strengths and superlattice amplitudes (lower left corner) as expected for a conventional condensate. In contrast to this, there is no sizable fringe visibility in the upper-right region, indicating that the localized phase is incoherent and has to be distinguished from the genuine Bose-Einstein condensate. This also applies to the narrow stripe along the $V_{ab}/J=0$ axis, where the fringe visibility exhibits an intermediate value. Again the groundstate is dominated by number states that have many bosonic particles at the deepest superlattice wells. In contrast to the area at the top right, not all of the particles are localized and thus many more number states contribute. The more number states contribute to the groundstate, the more the phase coherence is restored leading to more pronounced interference fringes.

In a recent experiment, G{\"u}nter et al. \cite{Kenneth} found that the successive admixture of fermions to a Bose-Einstein condensate in an optical lattice leads to a reduction of the interference fringes of the bosonic component. A detailed inspection of the visibility of the interference fringes $\nu$ in figure \ref{COND2}(b) at $\Delta/J=0$ might confirm their observation. With increasing interaction strength $V_{ac}$ the visibility also decreases rapidly. Albeit the boson-fermion interaction strength is fixed and the ratio between the particle numbers, $N_a/N_c$, is changed in the experiment, the effect on the bosons should be similar to the setup presented here.

\section{Basis truncation for Bose-Fermi mixtures}\label{secBFTRUNC}

The basis truncation scheme that was introduced in section \ref{truncationscheme}, can also be applied to Bose-Fermi mixtures. Before going to larger systems, we repeat the calculations of the previous section for a $I=10$, $N_a=N_c=5$  boson-fermion mixture using a truncated basis. The truncation of the basis again makes use of the importance criterion (\ref{TRUNC}). For the present example the basis dimension is reduced to $D_a \cdot D_c=72762$, thus about 15\% of the number states of the complete basis are kept. We use this rather large set because the phase diagram is not partitioned into several areas with different optimized bases --- for simplicity, only one basis optimized for $\Delta=V_{aa}=V_{ac}$ is employed. Of course, one can also work with an optimized basis for each point or for particular regions of the phase diagram. 

\begin{figure}
\includegraphics[width=1\textwidth]{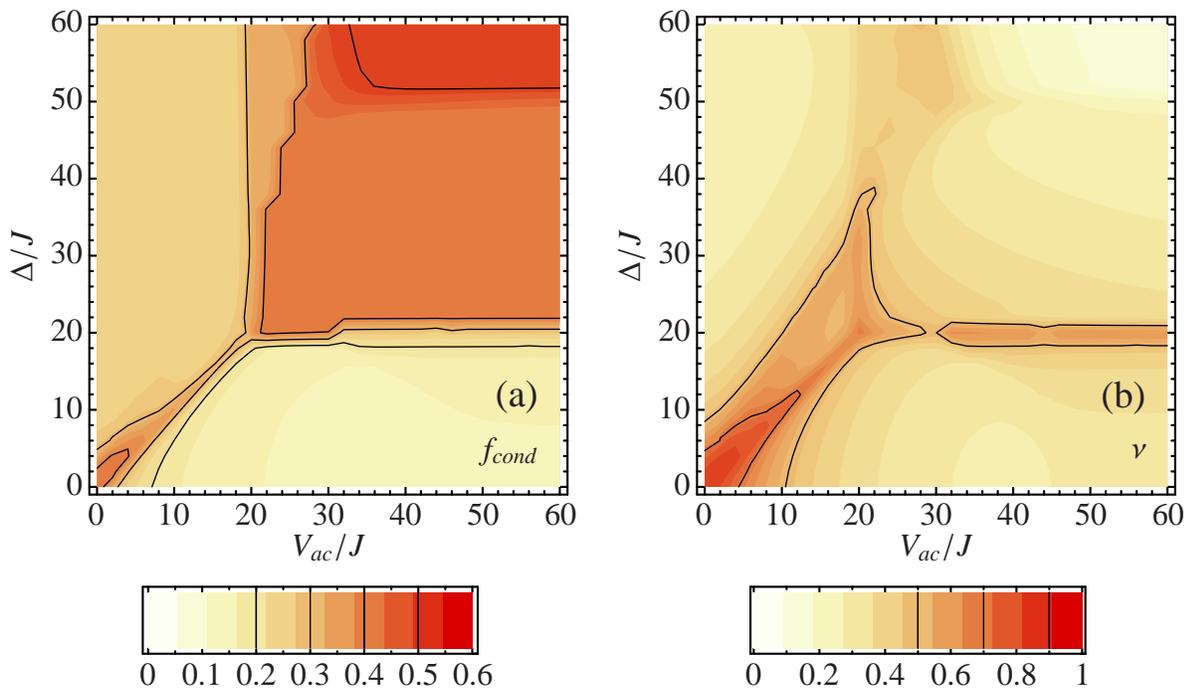}
\caption{Binary Bose-Fermi mixture with $N_a=N_c=5$ and $I=10$ for fixed $V_{aa}/J=20$ calculated with a truncated basis of roughly 15\% of the full dimension. (a) condensate fraction $f_{cond}$ of the bosonic species. (b) fringe visibility $\nu$ of the bosonic species. The comparison with figure \ref{COND2} illustrates the quality of the truncated calculation.\label{COND3}}
\end{figure}

For brevity, we only consider the condensate fraction and the fringe visibility, as shown in figure \ref{COND3}, for the truncated calculations. The general structure of the phase diagram of figure \ref{COND2} is well-reproduced with the truncated basis. The absolute value of the condensate fraction is underestimated in the truncated calculation as was discussed earlier. In close connection to this, the narrow stripe at the top left of figure \ref{COND3}(b) shows a significantly smaller visibility $\nu$ than in the complete calculation. This is  because number states with many particles at the same lattice sites are discarded by the truncation scheme due to the optimization to a region, where also the boson-fermion interaction plays a significant role. Although the coefficients $C_{\alpha\beta}$ of the missing number states would have been small, their contribution to the off-diagonal elements of the one-body density matrix is sizable. Their omission causes a reduction of coherence and thus the reduced visibility of the interference fringes. 

Despite the quantitative deviations in some special regions in the phase-diagram, its global structure can be fully described by the truncated basis. Naturally, the use of adapted bases in the different characteristic regions in the phase diagram would improve the quantitative agreement.

\section{Bose-Fermi mixtures beyond half-filling}

Finally, we investigate a system with $N_a=N_c=7$ particles and $I=10$ lattice sites. The complete basis for this system consists of $D_a \cdot D_c=1372800$ number states. For the presented calculations a truncation to 22\% of the complete basis is used.

\begin{figure}
\includegraphics[width=1\textwidth]{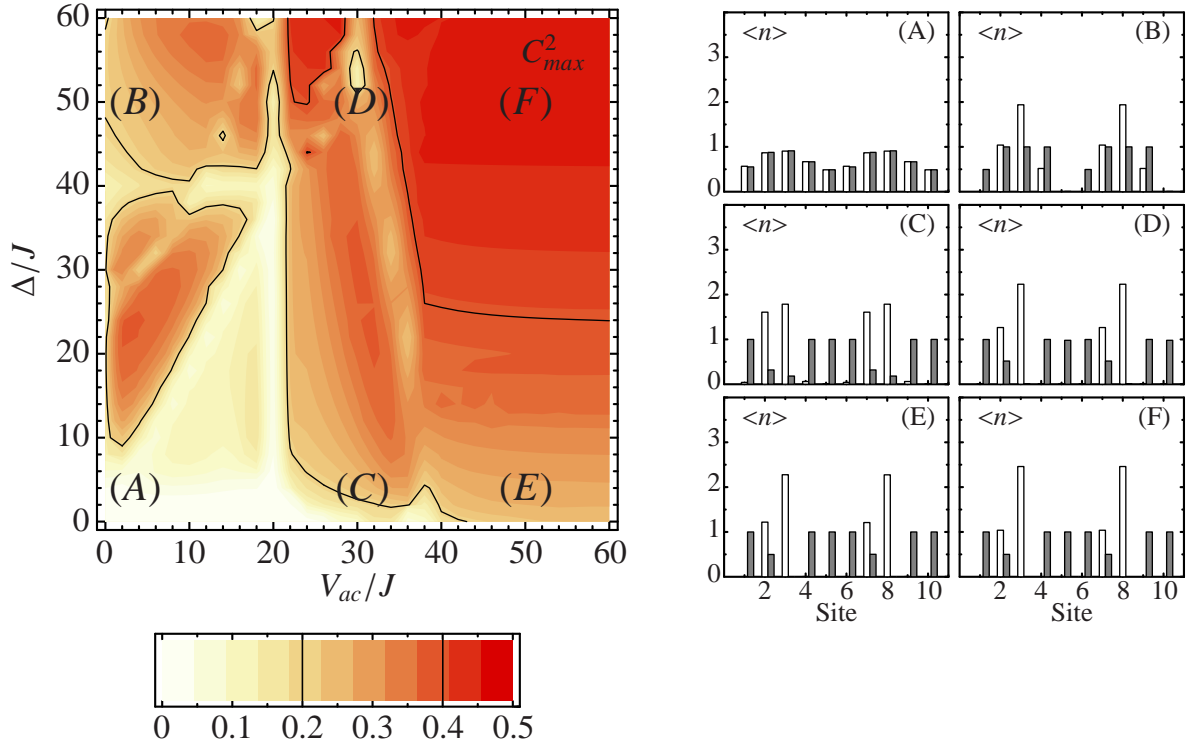}
\caption{Boson-fermion mixture with $I=10$, $N_a=N_c=7$. Left panel: maximum coefficient $C^2_{max}$ in the expansion of number states, right panel: mean occupation numbers  $n^{(a)}_{l}$ ($n^{(c)}_{l}$) of the bosonic (fermionic) species. \label{COND4} The boson-boson interaction energy is fixed at $V_{aa}/J=20$.}
\end{figure}

The maximum coefficient $C^2_{max}$, depicted in figure \ref{COND4} as a function of $V_{ac}/J$ and $\Delta/J$ for fixed $V_{aa}/J=20$, reveals a phase diagram with a structure different from the half-filling case shown in figure \ref{CMAX2}. The most significant difference is the absence of the Mott-insulator-like phase in the lower part of the phase diagram. Even in the limit of very strong repulsive interactions, a homogeneous Mott-insulator phase is not possible, since there are always lattice sites occupied with more than one particle due to the incommensurate filling. For this reason in region (A), where the boson-fermion interaction is negligible compared to the boson-boson repulsion, pairs of bosons and fermions occupy the lattice with a distribution following the topology of the superlattice. An increase of the superlattice amplitude (region (B)) leads to a bosonic double occupancy of the deepest superlattice wells, whereas the fermions are forced to spread over several sites with mean occupation numbers less than or equal to one. The horizontal line, which appeared in the half-filling case for superlattice amplitudes $\Delta$ equal to the boson-boson interaction strength $V_{aa}$, vanishes because of the incommensurability. For strong boson-fermion repulsion, any slight irregularity of the lattice---i.e. a small nonzero superlattice amplitude---leads to multiple boson occupancies at the deeper lattice sites. A comparison of the mean occupation numbers at (C) and (D) or at (E) and (F) shows that if the boson-fermion interaction exceeds the boson-boson interaction, the mean occupation numbers are stable against increasing superlattice amplitude. However, at the same time the increasing value of $C^2_{max}$ indicates that the number fluctuations decrease successively.

\section{Conclusions}

We have studied the phase diagram of binary boson-fermion mixtures in optical superlattices. For the symmetric half-filled system we identified a homogenous Mott-insulator-like phase at small superlattice amplitudes and strong repulsive interactions within the bosonic species and between the bosonic and the fermionic component. A phase of complete localization of the bosonic species can be realized already for moderate superlattice amplitudes if the boson-fermion repulsion clearly exceeds the boson-boson repulsion. These ``localized condensates'' are of a different nature than the genuine Bose-Einstein condensates in optical lattices as indicated by the absence of matter-wave interference fringes.

The phase-diagram of a boson-fermion mixture beyond half-filling is less structured, since the homogeneous Mott-insulator phase at small superlattice amplitudes is absent. In the region of strong boson-fermion interaction any slight irregularity within the lattice immediately leads to double occupancies of the deeper site. For increasing superlattice amplitude, the mean occupation numbers are almost unaltered, only the number fluctuations decrease.

Part of the calculations were performed using a controlled basis truncation scheme tailored to select the physically important basis states out of the complete set. This versatile tool allows to increase the particle numbers and lattice sizes accessible to diagonalization techniques. By explicit comparison with full calculations, we have shown that truncated bases containing only about one percent of the number states, already provide quantitatively reliable results. The simplicity of the basis truncation facilitates time-dependent calculations for systems of moderate size, as will be discussed in a separate publication \cite{markus}.
%
\section*{References}
\end{document}